\documentclass[10pt, superscriptaddress, reprint, amsmath, amssymb, aps, pra, showpacs]{revtex4-2}

\usepackage{amsmath,amssymb,mathrsfs}
\usepackage{cases}
\usepackage{graphicx}
\usepackage{dcolumn}
\usepackage{bm}
\usepackage{color, xcolor}
\usepackage{amsfonts,latexsym}
\usepackage{enumerate}
\usepackage{rotating}
\usepackage{epstopdf}
\usepackage{braket}
\usepackage{MnSymbol,bbding,pifont}

\usepackage{hyperref}
\hypersetup{
	colorlinks=true,
	linkcolor=blue,
	filecolor=blue,
	urlcolor=blue,
	citecolor=magenta
}

\newcommand{\Eq}[1]{Eq.~\eqref{#1}}
\newcommand{\eq}[1]{\eqref{#1}}
\newcommand{\Fig}[1]{Fig.~\ref{#1}}

\newcommand{\beq}{\begin{equation}}
	\newcommand{\eeq}{\end{equation}}
\newcommand{\beqa}{\begin{eqnarray}}
	\newcommand{\eeqa}{\end{eqnarray}}
\newcommand{\Beqa}{\begin{eqnarray*}}
	\newcommand{\Eeqa}{\end{eqnarray*}}

\def\bal#1\eal{\begin{align}#1\end{align}}
\def\Bal#1\Eal{\begin{align*}#1\end{align*}}

\newcommand{\me}{\mathrm{e}}

\newcommand{\overbar}[1]{\mkern 1.5mu\overline{\mkern-1.5mu#1\mkern-1.5mu}\mkern 1.5mu}

\begin{document}		
	
	\title{Parallel compression algorithm for fast preparation of defect-free atom arrays} 

	\author{Shangguo Zhu}	
	\author{Yun Long}
	\author{Mingbo Pu}	
	
	\affiliation{State Key Laboratory of Optical Technologies on Nano-Fabrication and Micro-Engineering, Institute of Optics and Electronics, Chinese Academy of Sciences, Chengdu 610209, China} 
	\affiliation{Research Center on Vector Optical Fields, Institute of Optics and Electronics, Chinese Academy of Sciences, Chengdu 610209, China} 
	\affiliation{School of Optoelectronics, University of Chinese Academy of Sciences, Beijing 100049, China}

	\author{Xiangang Luo}
	\email[]{lxg@ioe.ac.cn}
	
	\affiliation{State Key Laboratory of Optical Technologies on Nano-Fabrication and Micro-Engineering, Institute of Optics and Electronics, Chinese Academy of Sciences, Chengdu 610209, China} 
	\affiliation{School of Optoelectronics, University of Chinese Academy of Sciences, Beijing 100049, China} 
	
	\date{\today}
	
\begin{abstract}

Defect-free atom arrays have emerged as a powerful and versatile platform for quantum sciences and technologies, offering high programmability and promising scalability. The arrays can be prepared by rearranging atoms from a partially loaded initial array to the designated target sites. However, achieving large defect-free arrays presents challenges due to atom loss during rearrangement and the vacuum-limited lifetime which is inversely proportional to the array size. Efficient rearrangement algorithms which minimize time cost and atom loss are crucial for successful atom rearrangement. Here we propose a novel parallel compression algorithm which leverages multiple mobile tweezers to transfer atoms simultaneously. The total time cost could be reduced to scale linearly with the number of target sites. This algorithm can be readily implemented in current experimental setups.  

\end{abstract} 

\maketitle

\section{Introduction}
\label{sect:intro}

Since their invention in the 1970s, optical tweezers have evolved into sophisticated instruments for trapping and manipulating a diverse range of objects, including biological cells, bacteria, micro- and nano-particles, as well as ultracold atoms and molecules~\cite{Ashkin1986, Molloy2002, Zhang2008, Jones2015, Pesce2020, Liu2020, Liu2021, Zhang2022, Kaufman2021}. 
In the past decade, arrays of individual neutral atoms trapped in optical tweezers have emerged as a prominent and versatile platform, offering high programmability and promising scalability for quantum sciences and technologies.
This platform provides distinct advantages, including precise control and readout at the single-site level, flexible geometric configurations, and low entropy~\cite{Kaufman2021}.  
These exceptional features have contributed to numerous recent breakthroughs, including quantum simulations of exotic quantum matters~\cite{Browaeys2020, Kaufman2021, Ebadi2021, Scholl2021, Semeghini2021, Spar2022, Yan2022}, high-fidelity entanglement and quantum logic operations in quantum computations~\cite{Levine2019, Madjarov2020, Jenkins2022, Ma2022, Graham2022, Bluvstein2022}, and enhanced coherence, control and readout in quantum metrology~\cite{Madjarov2019, Norcia2019, Young2020}.

In contrast to the top-down approach of loading many cold atoms into optical lattices, atom arrays are constructed in a bottom-up manner by assembling individual atoms trapped in optical tweezers. 
A critical stage in the process involves achieving a full defect-free atom array through atom rearrangement, starting from an initial atom array which is partially stochastically loaded.  
Atom arrangement could be performed by dynamically changing the trap pattern generated by a spatial light modulator (SLM)~\cite{Kim2016, Lee2016}. 
However, limited by the speed of SLMs, this method is only applied to small-size atom arrays. 
Another strategy is to use acousto-optic deflectors (AODs) which generate extra mobile tweezers to rearrange atoms in the static tweezers produced by an SLM~\cite{Barredo2016, Endres2016, Barredo2018}. 
The atoms can be captured from or released to the static tweezers by ramping up or down the trap depths of the mobile tweezers, respectively. 
The position of mobile tweezers can be dynamically controlled by changing the tone frequencies of AODs. 
Then the atoms are transferred to the target sites via collision-free paths, which avoid atom loss due to collisions with other atoms. 
The application of this mobile tweezer strategy holds great promise for realizing large-size defect-free atom arrays.

Due to collisions with the residual gas, the vacuum-limited lifetime of a single atom trapped in an optical tweezer ranges from tens of seconds~\cite{Barredo2016, Schymik2020, Sheng2021, Ebadi2021} to approximately $23$ minutes~\cite{Manetsch2024} in a room-temperature apparatus. 
The vacuum-limited lifetime of an atom array is inversely proportional to the number of atoms. 
Especially for large array sizes, the atom rearrangement time should be less than the lifetime of the array. 
In the optimal case, the atom rearrangement time should only take a small portion of the vacuum-limited lifetime so that there will be sufficient time for subsequent operations of quantum simulation or computations.

On one hand, to achieve large-size defect-free atom arrays, one can increase the vacuum-limited lifetime by improving the experimental setups and techniques. 
Recently, the single-atom lifetime of over $6000~{\rm s}$ was reported for a system placed in a cryogenic environment~\cite{Schymik2021}, and later a defect-free array of over 300 atoms was achieved~\cite{Schymik2022}. 
On the other hand, one can minimize the atom rearrangement time by using optimized atom rearrangement algorithms. 
If we only consider transferring one atom at one time by a single mobile tweezer, several algorithms were proposed, including the heuristic shortest-moves-first algorithm~\cite{Barredo2016, Schymik2020}, Hungarian matching algorithm~\cite{Lee2017}, the compression algorithm~\cite{Schymik2020}, the LSAP algorithm~\cite{Schymik2020}, the heuristic cluster algorithm~\cite{Sheng2021}, and the heuristic heteronuclear algorithm for mixed-species atom arrays~\cite{Sheng2022, Singh2022}.
For these single-tweezer algorithms, the number of moves is at least the number of vacancies in the target sites. 
This limits the number of moves to scale linearly with the number of target sites $N$ at best. 
Recently, a pioneering method of simultaneously using multiple tweezers was demonstrated~\cite{Ebadi2021}. 
Later, more multiple-tweezer algorithms were proposed~\cite{Tian2023, Wang2023}, and the number of moves can be improved to scale as $N^{1/2}$.

In this work, we propose a novel heuristic algorithm called Parallel Compression Algorithm (PCA), which leverages multiple mobile tweezers to transfer atoms simultaneously to their target sites. 
This approach offers significant advantages by effectively reducing both the total number of moves required and the overall travel distance of the atoms during the preparation of defect-free atom arrays.
As a result, the total time for atom rearrangement can be reduced to scale at best linearly with the number of target sites $N$. 
Consequently, this approach enhances the probability of successfully achieving defect-free atom arrays. 
Moreover, the PCA is straightforward to implement and exhibits a short computation time, making it practical for implementation in existing experimental setups for atom arrays.

\section{Description of the problem} 
\label{sect:problem}

Initially, the laser-cooled atoms are stochastically loaded into the optical tweezer array, with a filling fraction denoted by $p$. 
After the atoms are ejected in pairs, we obtain a stochastic sub-Poissonian loading~\cite{Schlosser2001}. 
Each trap in the array has a probability of $p$ to be occupied by a single atom and a probability of $1-p$ to remain empty. 
Typically, the value of $p$ is around $50\%$~\cite{Schlosser2001}. 
By employing gray-molasses loading techniques
~\cite{Grunzweig2010, Lester2015, Brown2019, Jenkins2022, Aliyu2021, Angonga2022}, it is possible to enhance $p$ to exceed $90\%$. 
For generality, we assume $p = 50\%$ in the subsequent discussions. 
The goal is to achieve a full defect-free array of single atoms on the $N$ target traps by executing a deterministic sequence of operations to rearrange the atoms.

Given that the vacuum-limited lifetime of an atom array scales as $1/N$, it is imperative to devise an atom rearrangement algorithm that minimizes time cost. 
This becomes particularly crucial for large-scale quantum simulators or computers as $N$ increases. 
By minimizing the time required for atom rearrangement, sufficient time can be allocated for subsequent quantum simulation or computation operations. 
Additionally, reducing the total rearrangement time will enhance the success probability by reducing background gas collisions.

Using the mobile tweezers, atoms are captured from the source traps, transferred at a constant speed, and subsequently released to the target traps. 
During the capturing (releasing) process, the trap depth of the mobile tweezers are ramped up (down) in a time period $t_1$. 
Typically, $t_1 $ is about $ 15~ \rm{\mu s}$ to $ 60~ \rm{\mu s}$~\cite{Ebadi2021, Tian2023}. 
Let $t_2 = l / v$ be the time cost of moving an atom between two adjacent traps, where $v$ is the speed of the mobile tweezer and $l$ the distance between the adjacent static tweezers. 
Typically, the speed $v$ is about $75~ \rm{\mu m}/\rm{ms}$ to $ 130~ \rm{\mu m}/\rm{ms}$~\cite{Ebadi2021, Tian2023}. 
Then, for $l = 2~ \rm{\mu m}$, $t_2$ is about $15~ \rm{\mu s}$ to $27~ \rm{\mu s}$.
Overall, we can express the total rearrangement time $T$ as 
\begin{equation}
	\label{time}
	T = (C + R) t_1 + D t_2. 
\end{equation}
Here $C$ and $R$ are the number of capturing and releasing processes, respectively. 
$D $ is the total travel distance of all atoms in unit of $l$.

To minimize the overall rearrangement time, it is essential to minimize three key factors: $C$, $R$, and $D$. 
One challenge in the process is the imperfect transfer of atoms between the mobile and static tweezers. 
Each move of the atoms carries a success probability that is less than one. 
Consequently, reducing the values of $C$ and $R$ not only decreases the time spent on capturing and releasing atoms but also increases the overall success probability of the rearrangement process.

To prevent atom loss due to collisions or disturbances of the trapping potential, it is important to ensure that the paths of the mobile tweezers do not intersect with or come too close to the occupied traps. 
Consequently, if we opt to transfer atoms using paths between adjacent rows of traps, a relatively large spacing between the traps is required. 
To address this, we exclusively focus on collision-free paths that consist of lines connecting two neighboring traps and do not intersect with any occupied traps. 
By considering only these collision-free paths, we can ensure the safe and undisturbed transfer of atoms during the rearrangement process.

The choice of atom rearrangement algorithms, which determine how atoms are transferred from source traps to target traps, plays a crucial role in minimizing the overall time cost and maximizing the probability of achieving a defect-free atom array. 
However, exhaustively searching for the optimal rearrangement paths becomes challenging, particularly for large atom array sizes. 
In light of this, we focus on heuristic algorithms that offer an acceptable computation time while still providing effective solutions.

In summary, to obtain a superior heuristic rearrangement algorithm, it is essential to minimize (1) the total number of captures and releases $C+R$, (2) the overall travel distance $D$, and (3) the computation time required by the algorithm. 
By addressing these factors, we can optimize the efficiency and success probability of achieving a defect-free atom array.

\section{Parallel compression algorithm}
\label{sect:pca}

\begin{figure*}[ht!]
	\includegraphics[width = 0.95\linewidth]{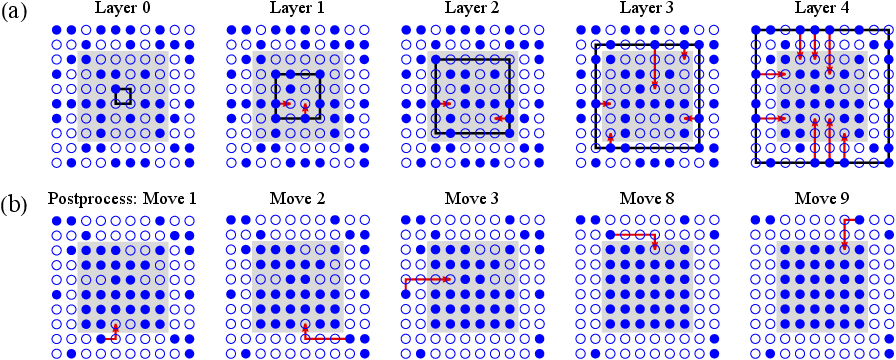}
	\caption{ Demonstration of the Parallel Compression Algorithm for a $6\times6$ target array (gray shaded region). (a) The parallel compression stage. The whole array is identified as several source layers (indicated by the black squares). Starting from Layer 1, the movable atoms on each side are transferred to the target traps in parallel with paths indicated by the red arrows. The blue dots and circles represent filled and empty traps, respectively. (b) The postprocess stage. The remaining empty traps are filled by moving the nearest available atoms. The moving paths are indicated by the read arrows with one turn. }
	\label{fig:graphDemo}
\end{figure*}

To maximize the efficiency of atom rearrangement under practical experimental conditions, we introduce the heuristic PCA. This algorithm employs the simultaneous generation of multiple mobile tweezers to enable parallel movement of atoms. 
The PCA consists of two main stages: (1) parallel compression and (2) postprocess.

\subsection{Parallel compression stage}
\label{sect:para}

In \Fig{fig:graphDemo}(a), we present a demonstration of the parallel compression stage for a $6\times 6$ target array. 
The process begins by partitioning the entire two-dimensional (2D) array into distinct layers. 
These layers are labeled as $0, 1, 2, 3, \cdots$ in an inward-to-outward manner. In the case of a square lattice, each layer forms a square shape.
Starting from Layer $1$, each layer is considered a source layer which provides movable atoms for the inner target region. 
These movable atoms are those which can be transferred along straight paths to the target traps within the inner target region. 
For convenience, the target region is divided into four triangular sections by the two diagonals. 
We require that the movable atoms from each side of the source layers are restricted to their corresponding triangular section and must not cross the diagonals.

To facilitate the parallel transfer of movable atoms, we employ a 2D AOD, which consists of a pair of AODs aligned along two orthogonal dimensions.
By setting the tone frequencies of one AOD, we create a 1D array of mobile tweezers which corresponds to the number of movable atoms on each side of the layer. 
Subsequently, the 1D array of multiple mobile tweezers can be transferred in parallel paths along a direction perpendicular to the array by modulating the tone frequency of the other AOD. 
Here using a single 2D AOD, we can address the four sides sequentially in a counterclockwise order.

Then, during the rearrangement process, the movable atoms are captured by ramping up the trapping potentials at the same time, then transferred in parallel straight paths, and released by ramping down the trapping potentials when reaching their respective target traps. 
The array of multiple mobile tweezers can be visualized as a shuttle bus which picks up a group of passengers and drops them off at different bus stops. 
This analogy captures the parallel nature of the atom transfer process and its resemblance to a transportation system. 
During a single transfer process, the number of capture operations remains constant at $1$, while the number of release operations corresponds to the number of distinct path lengths. 
The overall travel distance is determined by the longest path length.


Due to our parallel compression protocol, our algorithm has three main advantages compared to other multi-mobile-tweezer algorithms. 	
First, it avoids issues of intermodulation and limited mobility. In some other algorithms, mobile tweezers are constrained to move along the array of mobile tweezers. 
If many tweezers are generated, there will be intermodulation when driving multiple frequency tones, leading to a lower success rate~\cite{Tian2023, Wang2023}. 
Additionally, the mobility range of each frequency tone is inversely proportional to the number of tones produced by one AOD, limiting mobility in large arrays. 
This limitation is critical in neutral atom quantum computation, where arrays have reached 6100 atoms~\cite{Manetsch2024} and are approaching $10^4$.	
Our algorithm, with movement perpendicular to the array of mobile tweezers, fixes multiple frequency tones in one AOD and controls movement by driving the frequency tone of the other AOD, thus avoiding intermodulation and limited mobility issues. 
Second, our algorithm can be readily upgraded by using four 2D AODs to handle all four sides simultaneously, achieving a 4 times speed-up through increased parallelism. 	
Third, it can be further enhanced by integrating the stroboscopic technique, which allows generating many mobile tweezers simultaneously, each controlled individually along independent paths, significantly reducing the effective number of moves. 
The performance enhancement of the optimized algorithm, utilizing four 2D AODs and the stroboscopic technique is evaluated in Sect.~\ref{sect:performance}.


\subsection{Postprocess stage}

Following the completion of the parallel compression stage, a few empty traps may remain in the target array. 
To address this, a postprocess stage is performed, as illustrated in \Fig{fig:graphDemo}(b).

During the postprocess stage, the remaining empty traps are filled by relocating the nearest available atoms from outside the target array. 
The movement paths in this stage are no longer straight lines and often involve two successive straight paths with a single turn. 
Notably, the atoms are moved individually using a single mobile tweezer, without the parallelism employed in the earlier stage.

Here the idea is to maximize the transfer of atoms during the initial parallel compression stage, thereby capitalizing on the efficiency gains achieved enabled by parallelism. 
In Section \ref{sect:performance}, we will demonstrate that the contribution of the postprocess stage diminishes exponentially as we increase the number of reservoir atoms, which refer to the atoms initially present on the lattice.

\section{Performance of the Algorithm} 
\label{sect:performance}

\begin{figure}[ht!]
	\includegraphics[width=0.95\linewidth]{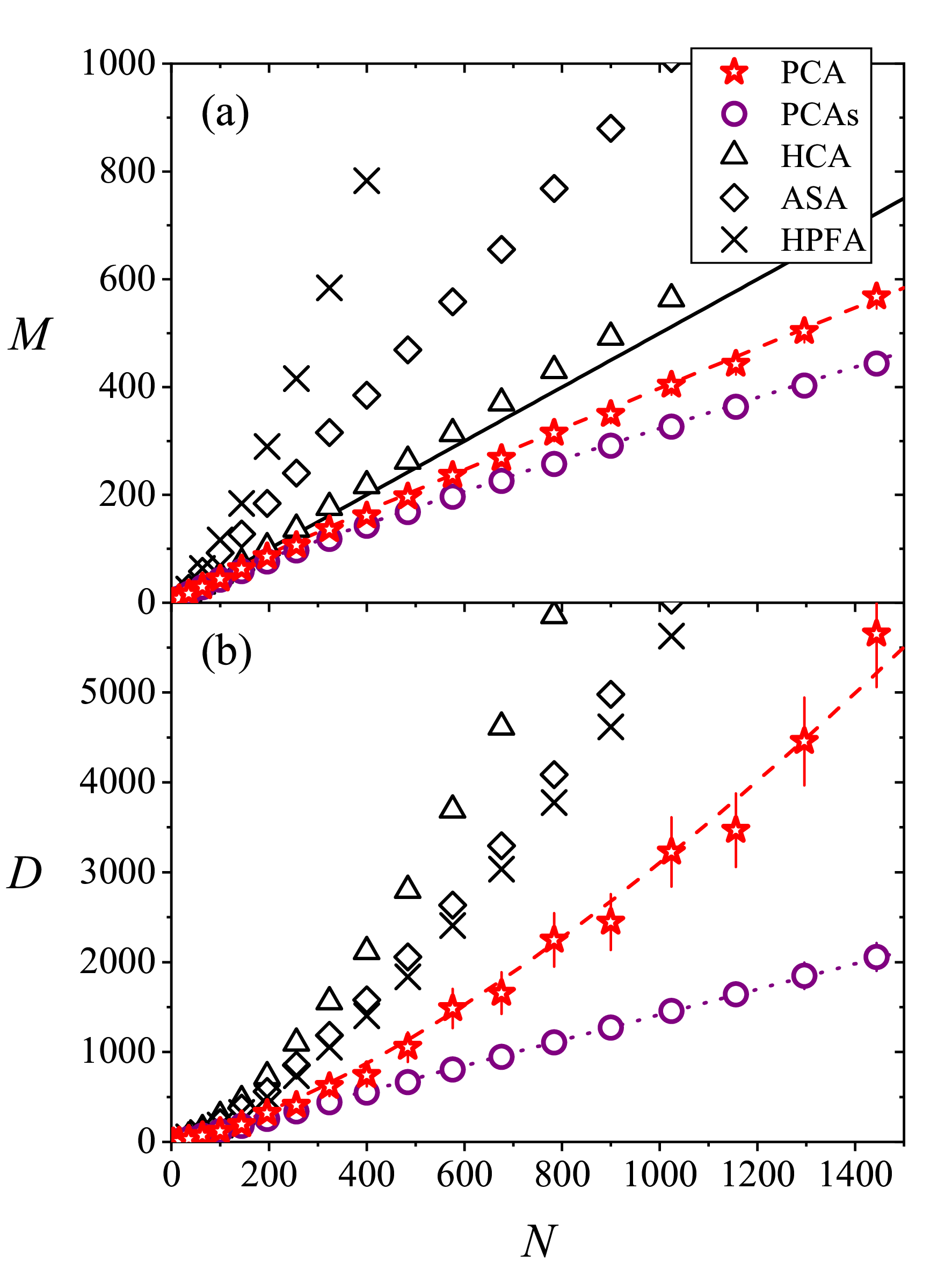}
	\caption{The number of effective moves $M$ in (a) and the total travel distance $D$ in (b) as functions of the number of target sites $N$ for different atom rearrangement algorithms, PCA (red stars), PCAs (purple circles), HCA (black triangles), ASA (black diamonds) and HPFA (black crosses). PCAs represents PCA with saturated reservoir atoms when the reservoir-to-target ratio $r \gtrsim 3$. The red dashed and purple dotted curves represent the corresponding fitting with functions shown in Eqs.~\eq{PCA} and \eq{PCAs}, respectively. In (a), the black solid line represents $N/2$. The PCA and PCAs data points are averaged over 10000 simulations and the error bars show the standard deviations. The HCA, ASA, and HPFA data points are taken from Ref.~\cite{Sheng2021}. }
	\label{fig:MD}
\end{figure}

Here we analyze the performance of the PCA. 
As showed in \Eq{time}, the time cost of atom rearrangement is determined by the number of capturing processes $C$, the number of releasing processes  $R$, and the total travel distances $D$. 
To compare with other algorithms, we define the number of effective moves $M$ as the average of the number of capturing and releasing processes,
\begin{equation}
	\label{m}
	M \equiv (C + R) / 2. 
\end{equation}
Starting from initial stochastically loaded arrays with filling fraction $p = 50\%$, we perform simulations of the PCA and study its performance for different target array sizes. 
For an $L \times L$ target array (the number of target sites $N = L^2$), we start with a larger $L'\times L'$ with $L' = \lceil p^{-1/2} L +1 \rceil $ to guarantee a sufficient number of reservoir atoms for most cases. 
Here $\lceil x \rceil$ is the ceiling function which gives the smallest integer greater than or equal to $x$.

In \Fig{fig:MD}, we plot $M$ and $D$ as functions of the number of target sites $N$ in comparison with other contemporary atom rearrangement algorithms. 
We see that compared to the heuristic cluster algorithm (HCA)~\cite{Sheng2021}, the A* searching algorithm (ASA)~\cite{Sheng2021}, and the heuristic path-finding algorithm (HPFA)~\cite{Barredo2016, Sheng2021}, the PCA has a significant improvement on performance in both $M$ and $D$. 
In \Fig{fig:MD}(a), the PCA is considerably below the $N/2$ limit for initial arrays with $50\%$ filling fraction. 
From fitting, we find the scaling functions in the parallel compression and post process stages, respectively,
\begin{align}
	\label{PCAParts}
	C_{\rm para} &\approx 2.887 N^{1/2}, ~ R_{\rm para} \approx 0.396 N, ~  D_{\rm para} \approx 0.60 N, \\
	C_{\rm post} &= R_{\rm post} \approx 0.154 N,\quad  D_{\rm post} \approx 0.079 N^{3/2}. 
\end{align}
As a result, we find the number of effective moves and the total travel distance
\begin{equation}
	\label{PCA}
	M \approx 0.35 N + 1.44 N^{1/2}, ~~ D \approx 0.079 N^{3/2} + 0.60 N. 
\end{equation}

The acceleration of the algorithm is primarily attributed to the significantly improved efficiency achieved through parallel transfers of atoms using multiple mobile tweezers during the parallel compression stage.
We see that $C_{\rm para}$ scales as $N^{1/2}$.
This is because during the parallel compression stage, the atoms are captured by multiple mobile tweezers in parallel, with the number of captures approximately equal to the total number of sides across all layers.
On the other hand, $R_{\rm para}$ scales as $N$ due to the possibility of multiple stops for releasing atoms. 
However, if the releasing process can be optimized such that the array of mobile tweezers does not stop while releasing atoms, $R_{\rm para}$ can be further improved to be equal to $C_{\rm para}$, resulting in a scaling behavior proportional to $N^{1/2}$. 
Regarding $D_{\rm para}$, it scales as $N$ as it is roughly the product of $C_{\rm para}$ and the average distance covered during a single parallel transfer (roughly $\propto N^{1/2}$).

In the postprocess stage, we see that both $C_{\rm post}$ and $R_{\rm post}$ scale as $N$, while $D_{\rm post}$ scales as $N^{3/2}$. 
Since the atoms are transferred to the remaining target sites one at a time, there is no acceleration through parallelism in this stage. 
As a result, the postprocess stage incurs a considerable time cost, especially for larger values of $N$, due to the higher scaling exponents.

Fortunately, the impact of the postprocess stage can be mitigated by increasing the size of the initial array. 
By selecting a large $L' = \lceil (3/p)^{1/2} L \rceil$, the contribution of the postprocess stage becomes negligible.
Consequently, the values of $M$ and $D$ approach their saturation values, as shown in \Fig{fig:MD}, indicated as PCAs.
From fitting, we find the saturation values approximately
\begin{equation}
	\label{PCAsParts}
	\overbar{C}_{\rm para} \approx 4.6 N^{1/2}, ~\overbar{R}_{\rm para} \approx 0.50 N, ~\overbar{D}_{\rm para} \approx 1.4 N.   
\end{equation}
Hence we obtain the saturation values of the number of effective moves $\overbar{M}$ and the total travel distance $\overbar{D}$, given by
\begin{equation}
	\label{PCAs}
	\overbar{M} \approx 0.25 N + 2.3 N^{1/2}, \qquad \overbar{D}\approx 1.4 N.
\end{equation}
We see that in the saturation case, $\overbar{M}$ and $\overbar{D}$ are determined by the parallel compression stage. 
Both $\overbar{M}$ and $\overbar{D}$ scale as $N$, at large $N$. 
From \Eq{time}, we find the approximate time cost formula
\begin{equation}
	\label{timePCAs}
	T \approx (0.5 t_1 + 1.4 t_2)N + 4.6 t_1 N^{1/2}. 
\end{equation}
We see that the total rearrangement time $T$ scales \textit{linearly} with $N$, similar to other multi-mobile-tweezer algorithms~\cite{Wang2023, Tian2023, Ebadi2021}.

As mentioned in Sect.~\ref{sect:para}, the algorithm can be readily upgraded to achieve a 4 times speed-up by using four 2D AODs to handle the four sides simultaneously.
In addition, by integrating the stroboscopic techniques, the mobile tweezers generated in the 2D AODs can be controlled individually. 
Consequently, the time cost of releasing processes becomes equal to that of capturing processes in the parallel compression stage, i.e., $\overbar{R}_{\rm para} = \overbar{C}_{\rm para} \approx 4.6 N^{1/2}$. 
Also, the number of effective moves becomes $\overbar{M} = 4.6N^{1/2}$. 
As a result, we find the time cost of the optimized algorithm 
\begin{equation}
	\label{timeOptimized}
	T^\ast = \frac{1}{4} ( 2 \overbar{C}_{\rm para} t_1 + \overbar{D}_{\rm para} t_2 )
	\approx 0.35 t_2 N + 2.3 t_1 N^{1/2}. 
\end{equation}
We see that the increased parallelism brought by multiple 2D AODs and stroboscopic techniques significantly reduces the overall time cost, resulting in a much smaller coefficient in the $N$-linear term compared to \Eq{timePCAs}.


\begin{figure}[ht!]
	\includegraphics[width=0.95\linewidth]{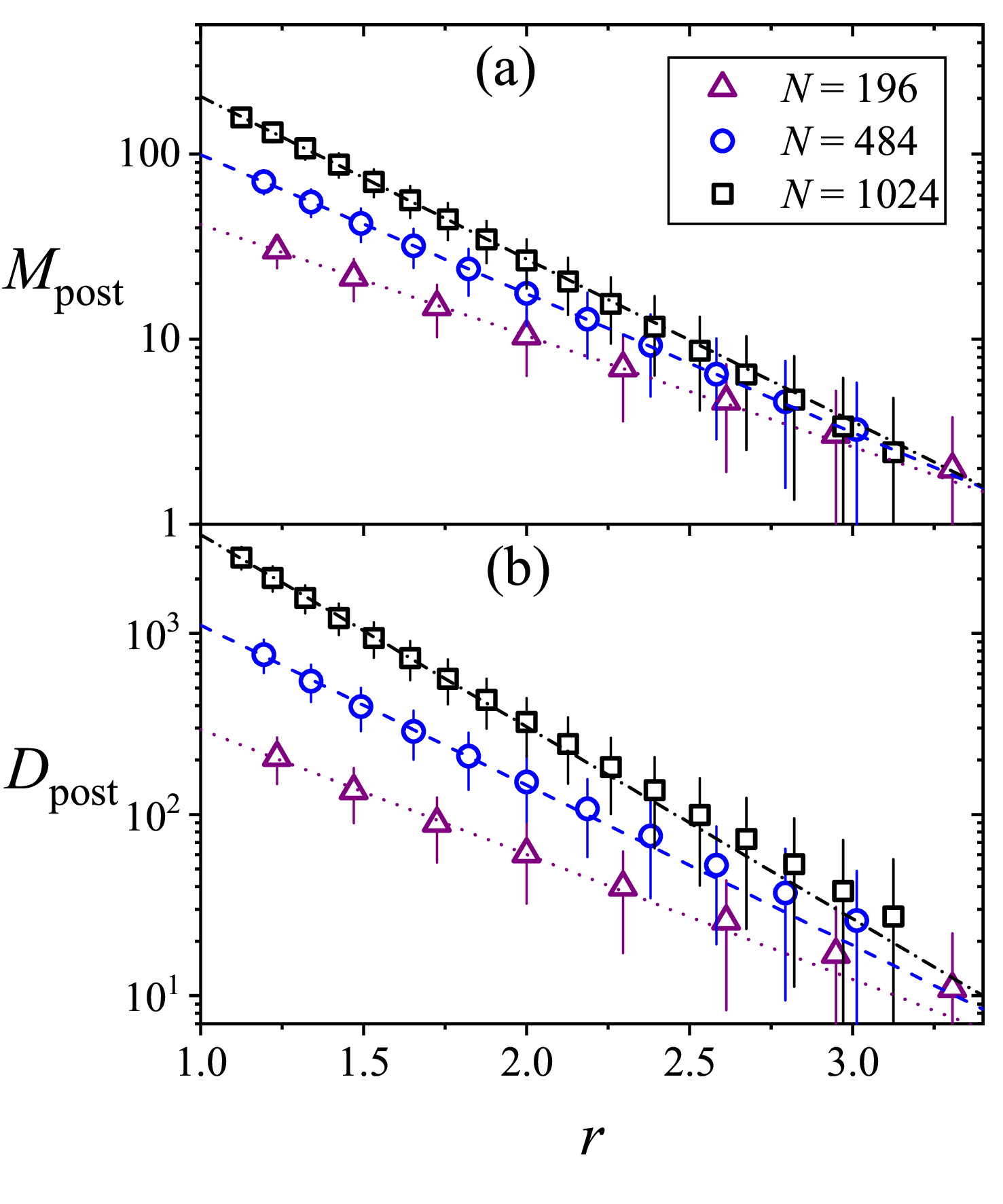}
	\caption{The number of moves $M_{\rm post}$ in (a) and the total travel distance $D_{\rm post}$ in (b) in the postprocess stage as functions of the reservoir-to-target ratio $r$, for the numbers of target sites $N = 196$ (purple triangles), $N = 484$ (blue circles), and $N = 1024$ (black squares). The purple dotted, blue dashed and black dash-dotted lines respectively represent the corresponding fitting with exponentially decaying functions $M_{\rm post} \approx 166.1 \me^{-1.384 r} $, $556.9 \me^{-1.727 r}$, $1544 \me^{-2.020 r}$ in (a), and $D_{\rm post} \approx 1439.1 \me^{-1.587 r} $, $8478.3 \me^{-2.034 r}$, $40205 \me^{-2.441 r}$ in (b). The points are averaged over 10000 simulations and the error bars show the standard deviations. } 
	\label{fig:MDpost}
\end{figure}

Further, we check how fast $M$ and $D$ approach their saturation values when we increase the size of the initial array. 
We perform simulations with fixed $L$ and varying $L'$. 
Here we consider values of $L'$ with the range $\lceil p^{-1/2} L +1 \rceil \le L' \le \lceil  (3/p)^{1/2} L \rceil$. 
In \Fig{fig:MDpost}, we plot the number of moves $M_{\rm post}$ and the total travel distance $D_{\rm post}$ in the postprocess stage as functions of the reservoir-to-target ratio $r \equiv pL'^2/L^2$, defined as the ratio between the average number of reservoir atoms and the number of target sites. 
We see that both $M_{\rm post}$ and $D_{\rm post}$ exhibit exponential decay as functions of $r$ when $r$ increases. 
The rate of decreasing is faster for larger target arrays (with greater $N$), indicating that the contribution from the postprocess stage decreases more rapidly as the target array size increases. 
Therefore, by appropriately increasing the number of reservoir atoms, we can significantly reduce the contribution from the postprocess stage. 
As a result, the vast majority of atom transfers are completed in the parallel compression stage, allowing us to fully leverage the acceleration provided by parallelism.

The implementation of the PCA algorithm is straightforward and can be easily integrated into a regular computer program. 
It requires a relatively short computation time, averaging around $1.8$ milliseconds, for a $14\times 14$ target array tested on a regular laptop computer with 8 CPU cores (Intel Core i7-1165G7) and 16 GB of RAM.

\section{Comparison with other multi-mobile-tweezer algorithms}
\label{sect:comparison}

The overall time cost of atom rearrangement using our PCA scales linearly with $N$, as shown in Eqs.\eq{timePCAs} and \eq{timeOptimized}. 
Other major multi-mobile-tweezer algorithms exhibit similar scaling. 
The Tetris algorithm scales as $\sim N^{0.9}$\cite{Wang2023}, while the PSCA~\cite{Tian2023} and the algorithm of Ref.~\cite{Ebadi2021} also demonstrate similar linear scaling with $N$.

The number of effective moves of the PCA scales as $N$ and can be improved to scale as $N^{1/2}$ for the optimized PCA. 
In comparison, the Tetris algorithm~\cite{Wang2023} scales as $\sim N$, the PSCA~\cite{Tian2023} as $\sim N^{0.4}$, and the algorithm in Ref.~\cite{Ebadi2021} as $\sim N^{0.5}$.

Now, we compare the overall time cost for a moderate array size of $N=400$. 
Assuming the time cost for capturing or releasing atoms is $t_1 = 35 \mu s$ and for moving one atom between two adjacent sites is $t_2= 30 \mu s$ (as set by the Tetris algorithm~\cite{Wang2023}), we find the overall time cost $T \approx 27 {\rm ms}$ and $T^\ast \approx 5.8 {\rm ms}$.
The Tetris algorithm reports a time cost of $T_{\rm Tetris}\approx 14 {\rm ms}$~\cite{Wang2023}. 
Since the other multi-mobile-tweezer algorithms do not provide explicit dependence of the time cost on $t_1$ and $t_2$, we need to set the same values of $t_1$ and $t_2$ for comparison. 
Setting $t_1 = 60 \mu s$ and $t_2 = 34.5 \mu s$ as in the PSCA algorithm~\cite{Tian2023}, we find $T \approx 37 {\rm ms}$ and $T^\ast \approx 7.6 {\rm ms}$. 
The PSCA algorithm~\cite{Tian2023} reports a time cost of $T_{\rm PSCA}\approx 14 {\rm ms}$, and the algorithm in Ref.\cite{Ebadi2021} gives $T_{\rm MTA1}\approx 58 {\rm ms}$\cite{Tian2023, Ebadi2021}.


We see that the $N$-linear time cost of our PCA is comparable to other multi-mobile-tweezer algorithms in performance, with the optimized PCA clearly outperforming them. 
Additionally, being free from intermodulation issues and limited mobility, our algorithm proves crucially valuable for atom rearrangement, especially for large arrays essential for fault-tolerant quantum computation~\cite{Bluvstein2024}.

\section{Summary}
\label{sect:summary}

In summary, we have introduced the Parallel Compression Algorithm, an efficient approach for rearranging atoms and achieving defect-free atom arrays from an initially partially stochastically loaded array. 
The key idea of this algorithm is to partition the array into layers and employ multiple mobile tweezers to transfer multiple movable atoms within each layer in parallel. 
This parallelism strategy allows for a significant reduction in the total time required, with a best-case scenario of scaling linearly with $N$, representing a substantial improvement over existing single-tweezer algorithms. 
Furthermore, the algorithm is straightforward to implement and exhibits fast computation times. 
It can also be readily upgraded by utilizing multiple 2D AODs and integrating the stroboscopic techniques, leading to a further acceleration.

The Parallel Compression Algorithm is poised to have significant and broad applications in quantum simulation and computation, particularly in the noisy intermediate-scale quantum era. 
By combining this algorithm with state-of-art experimental techniques such as gray-molasses loading to achieve higher filling fractions~\cite{Grunzweig2010, Lester2015, Brown2019, Jenkins2022, Aliyu2021, Angonga2022}, and employing cryogenic environments to extend the vacuum-limited lifetime~\cite{Schymik2021, Schymik2022}, it becomes possible to envision the realization of even larger defect-free atom arrays in forthcoming experiments. 
These advancements hold great promise for pushing the boundaries of quantum science and technology in the near future.

\begin{acknowledgments}
	We thank Wei Gou for carefully reading the manuscript and helpful comments. 
	S.Z. gratefully acknowledge support by the Youth Innovation Promotion Association, Chinese Academy of Sciences (2023399). 	
\end{acknowledgments}


\end{document}